\documentstyle[12pt,epsfig]{article}

\begin{document}
\begin {center}
{\bf {\Large
Hadron interferences in the proton induced coherent $\eta$ meson
production reaction on a scalar-isovector nucleus} }
\end {center}
\begin {center}
Swapan Das \footnote {Electronic address: swapand@barc.gov.in} \\
{\it Nuclear Physics Division,
Bhabha Atomic Research Centre  \\
Mumbai-400085, India }
\end {center}

\begin {abstract}
The coherent $\eta$ meson energy $E_\eta$ distribution spectra in the
proton nucleus reaction have been calculated to investigate the
$\pi^0 - \eta$ mesons' interference, in addition to the study of resonance
$N^*$ dynamics in the nucleus.
The
elementary reaction occurring in the nucleus is assumed to proceed as
$ pN \to pN^* $; $ N^* \to N\eta $. Born terms in the intermediate state
is also considered.
In
a scalar-isovector nucleus, this reaction occurrs because of $\pi^0$
and $\eta$ meson exchange interactions for the forward going proton and
$\eta$ meson; other meson exchange potentials do not contribute in this
process.
The
sensitivity of the cross section to the hadron nucleus interactions, and
the beam energy dependence of the cross section are studied for this
reaction.
\end {abstract}


PACS number(s): 25.40e, 13.30.Eg, 13.60.Le

\section{Introduction}

One of the current interests in the intermediate energy nuclear and
particle reactions is to explore the dynamics of $\eta$ meson which can
be produced either in (quasi)bound state or in continuum \cite{wgpr}.
Several
data sets for the $\eta$ meson production in the hadron induced reactions
are available from various laboratories, like COSY \cite{cosy} (see the
references there in), SATURNE \cite{satu}, Los Alamos \cite{loal},
Brookhaven \cite{broo}. The production of $\eta$ meson in heavy-ion
collisions was reported from GSI \cite{gsi}.
Due
to the advent of high duty electron accelerators at Jefferson Laboratory,
Bates, MAMI, ELSA, ..... etc, good quality data have been obtained for
the photo- and electro- production of $\eta$ meson  \cite{peeta}.
These
accelerator facilities, along with the newly developed sophisticated
detecting systems, provide ample scopes to investigate the physics of
$\eta$ meson.

The study of reaction mechanism for the $\eta$ meson production opens
various avenues to learn many exciting physics. Large and attractive
$\eta N$ scattering
length near the threshold production of this meson predicts the
existence of a new hadronic atom, i.e., (quasi) $\eta$-mesic nucleus
\cite{etmn, cosy2}.
The
$\pi^0-\eta$ mixing is shown to occur in the charge symmetry breaking
(CSB) reactions \cite{pemx}.
Being
an isoscalar particle, the $\eta$ meson can excite a nucleon to
$I=\frac{1}{2}$ resonances. Specifically, $N(1535)$ resonance,
$ I(J^P) = \frac{1}{2} (\frac{1}{2}^+) $, has large decay branching ratio
to $\eta$ meson and nucleon at the pole mass.
Therefore,
the $\eta$ meson production in the nuclear reaction is considered as a
potential tool to investigate the dynamics of $N(1535)$ in the nucleus.
Of course, this reaction can also be used to study the $\eta$ meson
nucleus interaction in the final state \cite{mosl, shrf}.

The $\eta$ meson can be produced through the hadronic interaction by
scattering off the pion or proton on the proton or nuclear target.
Theoretical
studies of these reactions, as done by various authors
\cite{petp, lawl, vett}, show that the $\eta$ meson in the final state
arises because of the decay of $N(1535)$ produced in the intermediate
state.
Sometime
back, Alvaredo and Oset \cite{oset} studied the coherent $\eta$ meson
production in the $(p,p^\prime)$ reaction on the spin-isospin saturated
nucleus: $ p + A(gs) \to p^\prime + A(gs) + \eta $.
The
elementary reaction in the nucleus is considered as
$ pN \to p^\prime N(1535) $; $ N(1535) \to N\eta $.
The
resonance $N(1535)$ in the intermediate state is produced due to the $\eta$
meson (a psuedoscalar-isoscalar meson) exchange interaction or potential
only, specifically, for the forward going proton and $\eta$ meson.
Contributions
from other meson exchange potentials, as discussed in Ref.~\cite{oset},
vanish for this reaction.

The importance of $N(1520)$ resonance in the $\eta$ meson production
reaction is discussed in Ref.~\cite{das14}. The earlier value of
the coupling constant $f_{\eta NN(1520)}$ was 6.72 where as the latest
value of it is 9.98. This coupling constant is much larger than the
$\eta N N(1535)$ coupling constant: $ g_{\eta NN(1535)} \simeq 1.86 $.
The
$\eta$ meson production cross section due to $N(1520)$ in the above
mentioned reaction is increased by a factor of $\sim 5$ because of the
enhancement in $f_{\eta NN(1520)}$.
As
shown in Fig.~2 of above reference, the $N^* \to N\eta$ decay probability
of $N(1520)$ resonance rises sharply over that of $N(1535)$ with the
increase in resonance mass. In fact, this probability for $N(1520)$ is
larger than that of $N(1535)$ at higher energy.
It
is also shown in Ref.~\cite{das14} that the contribution of $N(1535)$
resonance to the considered reaction is the largest at low energy, i.e.,
$\sim 1$ GeV, provided $f_{\eta NN(1520)}$ is taken equal to 6.72.
In
multi-GeV region, the distinctly dominant contribution to the reaction
(quoted above) arises due to $N(1520)$ resonance.
In
addition to these, the contributions to the cross section due to of
Born terms and other resonances (whose $N\eta$ branching ratio is $\geq 4\%$
\cite{pdg12}, i.e., $N(1650)$, $N(1710)$ and $N(1720)$ resonances) are
also presented in this work.

It should be mentioned that both $\pi$ and $\eta$ mesons are pseudoscalar
particles but $\pi^0$ exchange potential can't contribute to above
mentioned reaction since it is an isovector meson and the quoted reaction
involves isoscalar nucleus.
Contrast
to this, both $\pi^0$ and $\eta$ meson exchange interactions can contribute
to $p \to N^*$ excitation in the spin-saturated isovector nucleus.
Therefore, the coherent $\eta$ meson production in the $(p,p^\prime)$
reaction on the scalar-isovector nucleus can be used to study the
contribution of $\pi^0$ and $\eta$ meson exchange potentials (alongwith
their interference) to the reaction.
In
addition, this reaction can also be used to investigate the dynamics of
Born terms and nucleonic resonances, similar to those presented in
Ref.~\cite{das14}.

\begin{figure}[h]
\begin{center}
\centerline {\vbox {
\psfig{figure=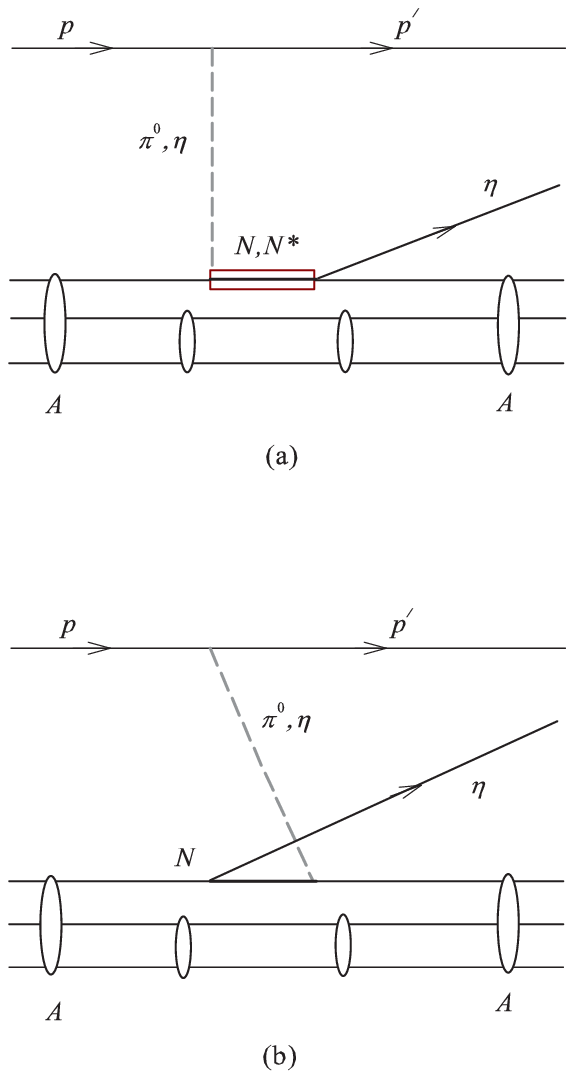,height=14.0 cm,width=6.0 cm}
}}
\caption{
(color online).
Schematic diagrams describing the mechanism of the considered reaction
(see text): (a) direct (or post-emission) and (b) cross (or pre-emission)
diagrams.
}
\label{fgdm}
\end{center}
\end{figure}

The diagrammatic presentation of the considered reaction is exhibited in
Fig.~\ref{fgdm}. The upper part of this figure, i.e., Fig.~\ref{fgdm}(a),
describes the direct or post-emission mechanism for the reaction
where
both Born term $N$ (presented by the bold solid line inside the
rectangle box) and resonance term $N^*$ (shown by the rectangle box)
have been incorporated.
The
lower part, i.e., Fig.~\ref{fgdm}(b), elucidates the cross or
post-emission mechanism for the reaction where only Born term $N$ is
considered, as the contribution of resonance $N^*$ term in this case
can be neglected compared that described in Fig.~\ref{fgdm}(a)
\cite{liu89}.

In the coherent meson production reaction, the meson in the final state,
i.e., the $\eta$ meson in the considered reaction, takes away almost
whole energy transferred to the nucleus,
i.e., $ E_\eta \approx (E_p-E_{p^\prime}) $, whereas the momentum of this
meson is adjusted by the recoiling nucleus. The state of the nucleus does
not change in this reaction.
In
the formalism for the coherent $\eta$ meson production in the proton
nucleus (scalar-isovector) reaction, the distorted wave functions of
protons and $\eta$ meson are expressed by the eikonal form.
The
optical potentials (appearing in $N^*$ propagator as well as in distorted
wave functions of protons) are worked out using $``t\varrho({\bf r})"$
approximation. The $\eta$ meson optical potential is evaluated following
that given in Ref.~\cite{oset}.
The
cross section for the coherent $\eta$ meson energy $E_\eta$ distribution
in the above reaction is calculated to study various aspects of it which
include (i) the resonance dynamics, (ii) the interference of $\pi^0$ and
$\eta$ meson exchange interactions, (iii) the sensitivity of the cross
section to the hadron nucleus interaction (optical potential), and (iv)
the beam energy dependence of the cross section.

\section{Formalism}

The Lagrangian ${\cal L}$ representing the meson baryon interaction
depends on their spin and parity. For the pseudoscalar $(0^-)$ meson
(i.e., $\pi$ or $\eta$ meson) coupling to the resonances in the
considered reaction, ${\cal L}$s are presented below
\cite{peters, blokland}.
For
the $\frac{1}{2}^+$ particle (i.e., $ N(940)$ or $N^* \equiv N(1710)$),
they can be expressed as
\begin{eqnarray}
{\cal L}_{\pi  NN~~} &=& -ig_\pi  F_\pi (q^2)
             {\bar N} \gamma_5 {\bf \tau} N \cdot {\bf \pi}  \nonumber \\
{\cal L}_{\eta NN~~} &=& -ig_\eta F_\eta (q^2)
                 {\bar N} \gamma_5 N \eta   \nonumber  \\
{\cal L}_{\pi  NN^*} &=& -ig^*_\pi  F^*_\pi (q^2)
             {\bar N^*} \gamma_5 {\bf \tau} N \cdot {\bf \pi} \nonumber \\ 
{\cal L}_{\eta NN^*} &=& -ig^*_\eta F^*_\eta (q^2)
                 {\bar N^*} \gamma_5 N \eta.  
\label{lag1}
\end{eqnarray}
$g_\pi$ ($\pi NN(940)$ coupling constant) $\simeq 13.4 $ \cite{erwe0},
$g_\eta$ ($\eta NN(940)$ coupling constant) $\simeq 7.93 $ \cite{chiang};
$g^*_\pi$ ($\pi NN(1710)$ coupling constant) $=1.2$,
and  $g^*_\eta$ ($\eta NN(1710)$ coupling constant) $\simeq 4.26 $.
For
$\frac{1}{2}^-$ resonance $N^*$, i.e., $N(1535)$ and $N(1650)$, the
forms for ${\cal L}$ are given by
\begin{eqnarray}
{\cal L}_{\pi NN^*} &=& -ig^*_\pi F^*_\pi (q^2)
            {\bar N^*} {\bf \tau} N \cdot {\bf \pi}  \nonumber \\
{\cal L}_{\eta NN^*} &=& -ig^*_\eta F^*_\eta (q^2) {\bar N^*} N \eta.
\label{lag2}
\end{eqnarray}
$ g^*_\pi \simeq 0.71 $, $ g^*_\eta \simeq 1.86 $ for $N(1535)$ and
$ g^*_\pi \simeq 0.83 $, $ g^*_\eta \simeq 0.67 $ for $N(1650)$.
For
$ N(1520) \frac{3}{2}^- $, ${\cal L}$s can be written as
\begin{eqnarray}
{\cal L}_{\pi NN^*} &=&  \frac{f^*_\pi}{m_\eta} F^*_\pi (q^2)
                        {\bar N^{*\mu}} \gamma_5 {\bf \tau} N
                         \cdot \partial_\mu {\bf \pi} \nonumber \\
{\cal L}_{\eta NN^*} &=& \frac{f^*_\eta}{m_\eta} F^*_\eta (q^2)
                        {\bar N^{*\mu}} \gamma_5 N \partial_\mu \eta.
\label{lag3}
\end{eqnarray}
$ f^*_\pi = 6.54 $ and $ f^*_\eta \simeq 9.98 $.
For
$\frac{3}{2}^+$ resonance \cite{das04}, i.e., $ N^* \equiv N(1720) $,
${\cal L}$s are given by
\begin{eqnarray}
{\cal L}_{\pi  NN^*} &=& \frac{f^*_\pi}{m_\eta} F^*_\pi (q^2)
                        {\bar N^{*\mu}} {\bf \tau} N
                         \cdot \partial_\mu {\bf \pi}   \nonumber  \\
{\cal L}_{\eta NN^*} &=& \frac{f^*_\eta}{m_\eta} F^*_\eta (q^2)
                        {\bar N^{*\mu}} N \partial_\mu \eta.
\label{lag4}
\end{eqnarray}
$ f^*_\pi \simeq 0.64 $ and $ f^*_\eta \simeq 1.15 $.
The coupling constants (i.e., $g^*$s and $f^*$s) are extracted from the
measured decay widths of the resonances,
i.e., $ N^* \to N\pi $ and $ N^* \to N\eta $ \cite{pdg12}.
$F_{\pi(\eta)}(q^2)$
and $F^*_{\pi(\eta)}(q^2)$ are $\pi(\eta) NN$ and $\pi(\eta) NN^*$
form factors at the respective vertices \cite{chiang}:
\begin{equation}
F_M(q^2)=F^*_M(q^2) = \frac{\Lambda^2_M - m^2_M}{\Lambda^2_M - q^2};
~~~~~ (M \equiv \pi^0, \eta).
\label{ffc}
\end{equation}
$ q^2 [ = q^2_0 - {\bf q}^2 ] $ in this equation is the four-momentum
transfer to the nucleus, i.e., $ q_0 = E_p - E_{p^\prime} $
and $ {\bf q} = {\bf k}_p - {\bf k}_{p^\prime} $.  The form factors are
normalized to unity when the mesons are in on-shell. Values of the
length parameters are $ \Lambda_\pi = 1.3 $ GeV and
$ \Lambda_\eta = 1.5 $ GeV \cite{chiang}.

The $T$-matrix $T_{fi}$ of the considered reaction can be written as
\begin{equation}
T_{fi} =  \sum_{M=\pi^0,\eta} [ T_B(M) + T_{N^*}(M) ],
\label{tmx1}
\end{equation}
where $T_B(M)$ represents the $T$-matrix due to Born terms arising because
of either $\pi^0$ or $\eta$ meson exchange potential. It is given
by
\begin{equation}
T_B(M) = \Gamma_{NN\eta} \Lambda_N (S)  V_M(q) 
\int d{\bf r} \chi^{(-)*} ({\bf k}_\eta, {\bf r}) G_N 
\varrho_I ({\bf r}) \chi^{(-)*} ({\bf k}_{p^\prime}, {\bf r})
\chi^{(+)} ({\bf k}_p, {\bf r}).
\label{tmxB}
\end{equation}
The factor $\varrho_I$ in above equation denotes the isospin averaged
density distribution of the nucleus. $\chi$s represent the wave functions
for the continuum particles. These quantities have been elaborated later.
$\Gamma_{NN\eta}$ describes the interaction for the emission of $\eta$
meson at $\eta NN$ vertex in the final state.
$V_M(q)$
represents the pseudoscalar meson (i.e., $M \equiv \pi^0 ~\mbox{or}~ \eta$)
exchange potential between the beam proton and a nucleon in the nucleus,
shown by the dash line in Fig.~\ref{fgdm}:
$ V_M(q) = \Gamma_{MNN} G_M (q^2) \Gamma_{Mpp^\prime} $.
In
this equation, $\Gamma_{MNN}$ denotes the interaction at $MNN$ vertex
(in the nucleus) where as $\Gamma_{Mpp^\prime}$ represents that at
$Mpp^\prime$ vertex (i.e., meson-projectile-ejectile vertex).
These $\Gamma$s are
addressed by $ {\cal L}_{\pi(\eta) NN} $ in Eq.~(\ref {lag1}).
$G_M(q^2)$
is the virtual $\pi^0(\eta)$ meson propagator, i.e.,
$ G_{\pi^0(\eta)} (q^2) = - \frac{1}{m^2_{\pi^0(\eta)} - q^2} $.
$\Lambda_N (S)$
represents the spin $S$ dependent part of the nucleon propagator, i.e.,
$\Lambda_N (S=\frac{1}{2})$ in Table-\ref {tbs}.
The
scalar part of the nucleon Born propagator $G_N$ for direct (D) and cross
(C) channels, as illustrated in Fig.~\ref{fgdm}, are given
by
\begin{equation}
G^D_N = \frac{1}{s-m^2_N}; ~~~~~ G^C_N = \frac{1}{u-m^2_N},
\label{nbpr}
\end{equation}
where $s$ and $u$ are invariant Mandelstam kinematical variables, see
page 28 in Ref.~\cite{erwe0}.

\begin{table}
\caption{ $\Lambda_{N(N^*)}(S)$ for spin $S=$ $\frac{1}{2}$ and
$\frac{3}{2}$ fermions. }
\centering
\begin{tabular} { c c }
\hline
Spin($S$)      &  $\Lambda(S)$    \\  [0.5ex]
\hline
$\frac{1}{2}$  &  $\{ \not k + m_{N^*} \} $  \\
$\frac{3}{2}$  &
$ \{ \not k + m_{N^*} \}
\left [ g^\mu_\nu  -\frac{\gamma^\mu \gamma_\nu}{3}
-\frac{ \gamma^\mu k_\nu - \gamma^\nu k_\mu }{ 3m_{N^*} }
-\frac{2k^\mu k_\nu}{3m^2_{N^*}}
\right  ]$   \\  [1ex]
\hline
\end{tabular} 
\label{tbs}
\end{table}

The resonance contribution to $T$-matrix, i.e., $T_{N^*}(M)$ in
Eq.~(\ref {tmx1}), is given by
\begin{equation}
T_{N^*}(M) = \sum_{N^*} \Gamma_{N^* \to N\eta}  \Lambda_{N^*} (S)
V_M(q)   
\int d{\bf r} \chi^{(-)*} ({\bf k}_\eta, {\bf r}) G_{N^*} 
\varrho_I ({\bf r}) \chi^{(-)*} ({\bf k}_{p^\prime}, {\bf r})
\chi^{(+)} ({\bf k}_p, {\bf r}).
\label{tmxR}
\end{equation}
$\varrho_I ({\bf r})$ and $\chi$s are also appeared in Eq.~(\ref{tmxB}).
$\Gamma_{N^* \to N\eta}$ denotes $N^* \to N\eta$ decay in the final
state. $V_M(q)$ in this case is given by
$ V_M(q) = \Gamma_{MNN^*} G_M(q^2) \Gamma_{Mpp^\prime} $,
where
$\Gamma_{M NN^*}$ represents the interaction at $MNN^*$ vertex
(in the nucleus) described by $ {\cal L}_{\pi(\eta) NN^*} $ in
Eqs.~(\ref {lag1} - \ref{lag4}). Other quantities in $V_M(q)$ are already
defined below Eq.~(\ref{tmxB}).
The
spin dependent part of $N^*$ propagator $\Lambda_{N^*} (S)$ is
expressed in Table-\ref {tbs}.
The
scalar part of this propagator $G_{N^*}$ in Eq.~(\ref {tmxR}), according
to Fig.~\ref{fgdm}(a), can be expressed as
\begin{equation}
G_{N^*} (m) = \frac{1}{m^2-m^2_{N^*} + im_{N^*}\Gamma_{N^*}(m)},
\label{nsp0}
\end{equation}
where $m_{N^*}$ are the pole mass of the resonance $N^*$. $m(\equiv s)$
is the invariant mass of $\eta$ meson and nucleon, arising due to the
decay of $N^*$.
Since
the cross or pre-emission channel in this case (as mentioned earlier)
can be neglected, it is not considered to evaluate $T$-matrix. Indeed,
this equation represents the resonance propagator in the free space as
the resonance nucleus interaction in it (which is considered later) is
omitted.

$\Gamma_{N^*}(m)$
in Eq.~(\ref{nspr}) represents the total width of $N^*$ for its mass equal
to $m$. The experimentally determined value of it at its pole mass,
i.e., $m=m_{N^*}$, for all considered resonances are listed in
Table-\ref{tbw}. Since a resonance $N^*$ can decay into various channels,
$\Gamma_{N^*}(m)$ consists of partial decay widths as written below
\cite{pdg12}.

For $ N^* \equiv N(1520) $ resonance,
\begin{equation}
\Gamma_{N^*}(m)
= \Gamma_{N^* \to N\pi}(m)|_{l=2} + \Gamma_{N^* \to \Delta \pi}(m)|_{l=0}
+ \Gamma_{N^* \to \Delta \pi}(m)|_{l=2} + \Gamma_{N^* \to N \eta}(m)|_{l=2},
\label{wd1520}
\end{equation}
where
$l$ is the angular momentum associated with the decay.
$ \Gamma_{N^* \to N\pi}(m)|_{l=2} \approx 0.65 \Gamma_{N^*}(m) $,
$ \Gamma_{N^* \to \Delta \pi}(m)|_{l=0} = 0.2  \Gamma_{N^*}(m) $,
$ \Gamma_{N^* \to \Delta \pi}(m)|_{l=2} = 0.15 \Gamma_{N^*}(m) $,
$ \Gamma_{N^* \to N \eta}(m)|_{l=2} = 2.3 \times 10^{-3}
\Gamma_{N^*}(m) $.

For $ N^* \equiv N(1535) $ resonance,
\begin{equation}
\Gamma_{N^*}(m)
= \Gamma_{N^* \to N\pi}(m)|_{l=0} + \Gamma_{N^* \to N \eta}(m)|_{l=0}
+ \Gamma_{N^* \to N \pi \pi}(m);
\label{wd1535}
\end{equation}
with
$ \Gamma_{N^* \to N \pi \pi}(m) = 0.1 \Gamma_{N^*}(m_{N^*}) $ \cite{fadlt}.
$ \Gamma_{N^* \to N\pi}(m)|_{l=0} = 0.48 \Gamma_{N^*}(m) $ and
$ \Gamma_{N^* \to N\eta}(m)|_{l=0} = 0.42 \Gamma_{N^*}(m) $.

For $ N^* \equiv N(1650) $ resonance,
\begin{equation}
\Gamma_{N^*}(m)
= \Gamma_{N^* \to N\pi}(m)|_{l=0} + \Gamma_{N^* \to \Delta\pi}(m)|_{l=2}
+ \Gamma_{N^* \to N\eta}(m)|_{l=0};
\label{wd1650}
\end{equation}
with
$ \Gamma_{N^* \to N\pi}(m)|_{l=0} = 0.75 \Gamma_{N^*}(m) $,
$ \Gamma_{N^* \to \Delta\pi}(m)|_{l=2} = 0.15 \Gamma_{N^*}(m) $
and
$ \Gamma_{N^* \to N\eta}(m)|_{l=0} = 0.1 \Gamma_{N^*}(m) $.

For $ N^* \equiv N(1710) $ resonance,
\begin{equation}
\Gamma_{N^*}(m)
= \Gamma_{N^* \to N\pi}(m)|_{l=1} +  \Gamma_{N^* \to \Delta \pi}(m)|_{l=1}
+ \Gamma_{N^* \to N\eta}(m)|_{l=1} + \Gamma_{N^* \to \Lambda K}(m)|_{l=1};
\label{wd1710}
\end{equation}
with
$ \Gamma_{N^* \to N\pi}(m)|_{l=1} = 0.2 \Gamma_{N^*}(m) $,
$ \Gamma_{N^* \to \Delta\pi}(m)|_{l=1} = 0.4 \Gamma_{N^*}(m) $,
$ \Gamma_{N^* \to N\eta}(m)|_{l=1}    = 0.3 \Gamma_{N^*}(m) $
and
$ \Gamma_{N^* \to \Lambda K}(m)|_{l=1} = 0.1 \Gamma_{N^*}(m) $.

For $ N^* \equiv N(1720) $ resonance,
\begin{equation}
\Gamma_{N^*}(m)
= \Gamma_{N^* \to N\pi}(m)|_{l=1} + \Gamma_{N^* \to \Delta\pi}(m)|_{l=1}
+ \Gamma_{N^* \to N\eta}(m)|_{l=1} + \Gamma_{N^* \to \Lambda K}(m)|_{l=1};
\label{wd1720}
\end{equation}
with
$ \Gamma_{N^* \to N\pi}(m)|_{l=1} = 0.11 \Gamma_{N^*}(m) $,
$ \Gamma_{N^* \to \Delta\pi}(m)|_{l=1} = 0.75 \Gamma_{N^*}(m) $,
$ \Gamma_{N^* \to N\eta}(m)|_{l=1} = 0.04 \Gamma_{N^*}(m) $
and
$ \Gamma_{N^* \to \Lambda K}(m)|_{l=1} = 0.1 \Gamma_{N^*}(m) $.

\begin{table}
\caption{ Resonance width $ \Gamma_{N^*} (m_{N^*}) $ at pole mass
$m_{N^*}$ in MeV \cite{pdg12}. }
\centering
\begin{tabular} { c c }
\hline
Resonance $N^*$ & $ \Gamma_{N^*} (m_{N^*}) $  \\  [0.5ex]
\hline
$N(1520)$       &    115                      \\
$N(1535)$       &    150                      \\
$N(1650)$       &    150                      \\
$N(1710)$       &    100                      \\
$N(1720)$       &    250                      \\  [1ex]
\hline
\end{tabular} 
\label{tbw}
\end{table}

The partial decay width of a resonance $N^*$ decaying to a baryon $B$
and a meson $M$, i.e., $\Gamma_{N^* \to BM} (m) |_l$, varies
with its mass $m$ \cite{manley} as
\begin{equation}
\Gamma_{N^* \to BM} (m) |_l = \Gamma_{N^* \to BM} (m_{N^*})
\left [ \frac{\Phi_l (m) }{\Phi_l (m_{N^*})} \right ].
\label{wdth}
\end{equation} 
The phase-space factor $\Phi_l (m)$ is given by
$ \Phi_l (m) = \frac{\tilde {k}}{m} B^2_l (\tilde {k}R) $,
where
$\tilde {k}$ is the relative momentum of the decay products
(i.e., $B$ and $M$) in their c.m. frame. $ B_l(\tilde {k}R) $ is 
Blatt-Weisskopf barrier-penetration factor, listed in Table \ref{tbb}.
$R$ ($=0.25$ fm) is the interaction radius.

\begin{table}
\caption{ Blatt-Weisskopf barrier-penetration factor $ B_l(\tilde {k}R) $
          \cite{manley}. }
\centering
\begin{tabular} { c c }
\hline
  $l$   &  $ B^2_l (x = \tilde{k}R) $  \\  [0.5ex]
\hline
   0    &       1                      \\
   1    &  $ x^2/(1+x^2)      $        \\
   2    &  $ x^4/(9+3x^2+x^4) $        \\  [1ex]
\hline          
\end{tabular} 
\label{tbb}
\end{table}

The five fold differential cross section of the considered reaction
can be written as
\begin{equation}
\frac{d\sigma}{dE_{p^\prime}d\Omega_{p^\prime}d\Omega_\eta}
=K_F <|T_{fi}|^2>,
\label{dcrss}
\end{equation}
where the annular brackets around $|T_{fi}|^2$ represent the average
over spins in the initial state and the summation over spins in the
final state. $K_F$ is the kinematical factor for the reaction:
\begin{equation}
K_F = \frac{\pi}{(2\pi)^6}
\frac{ m^2_p m_A k_{p^\prime} k^2_\eta }
{ k_p | k_\eta (E_i-E_{p^\prime}) - E_\eta {\bf q} . {\hat k}_\eta | }.
\label{kfc}
\end{equation}
All symbols carry their usual meanings.

\section{Results and Discussion}

The differential cross sections
$ \frac{ d\sigma }{ dE_{p^\prime} d\Omega_{p^\prime} d\Omega_\eta } $
have been calculated for the coherent $\eta$ meson energy $E_\eta$
distribution in the $(p,p^\prime)$ reaction on $^{14}$C, a
scalar-isovector nucleus.
To
describe the plane wave results, $\chi$s in $T$-matrices in Eqs~(\ref{tmxB})
and (\ref{tmxR}) are given by
$\chi^{(+)} ({\bf k}_p, {\bf r}) = e^{i {\bf k}_p. {\bf r} }$ for the beam
proton $p$, and
$\chi^{(-)^*} ({\bf k}_X, {\bf r}) = e^{-i {\bf k}_X. {\bf r} }$ for
a particle in the final state, i.e., $X$ is either ejectile proton
$p^\prime$ or $\eta$ meson. The resonance nucleus interaction or optical
potential $V_{ON^*}({\bf r})$ is not considered at this stage.
The
spatial density distribution $\varrho ({\bf r})$ of $^{14}$C nucleus,
as extracted from the electron scattering data \cite{andt}, is given by
\begin{eqnarray}
\varrho ({\bf r})
= \varrho_0 [1+w(r/c)^2] e^{-(r/c)^2};
~~~ w=1.38, ~c=1.73 ~\mbox{fm}.     
\label{vrrh}
\end{eqnarray}
This density distribution is normalized to the mass number of nucleus.

The isospin averaged nuclear density distribution $\varrho_I ({\bf r})$,
appearing in Eqs.~(\ref{tmxB}) and (\ref{tmxR}), is related to
$\varrho ({\bf r})$ as
\begin{equation}
\varrho_I ({\bf r})
= \left [ \frac{Z}{A} C_{is}(p) + \frac{(A-Z)}{A} C_{is}(n) \right]
  \varrho ({\bf r}),
\label{vrhI}
\end{equation}
where $C_{is}(p)$ and $C_{is}(n)$ are the isospin matrix elements
for the proton and neutron respectively. $C_{is}(p)=+1$ and $C_{is}(n)=-1$
are the values for $\pi^0$ meson exchange potential where as both of them
are equal to +1 for $\eta$ meson exchange potential.

\begin{figure}[h]
\begin{center}
\centerline {\vbox {
\psfig{figure=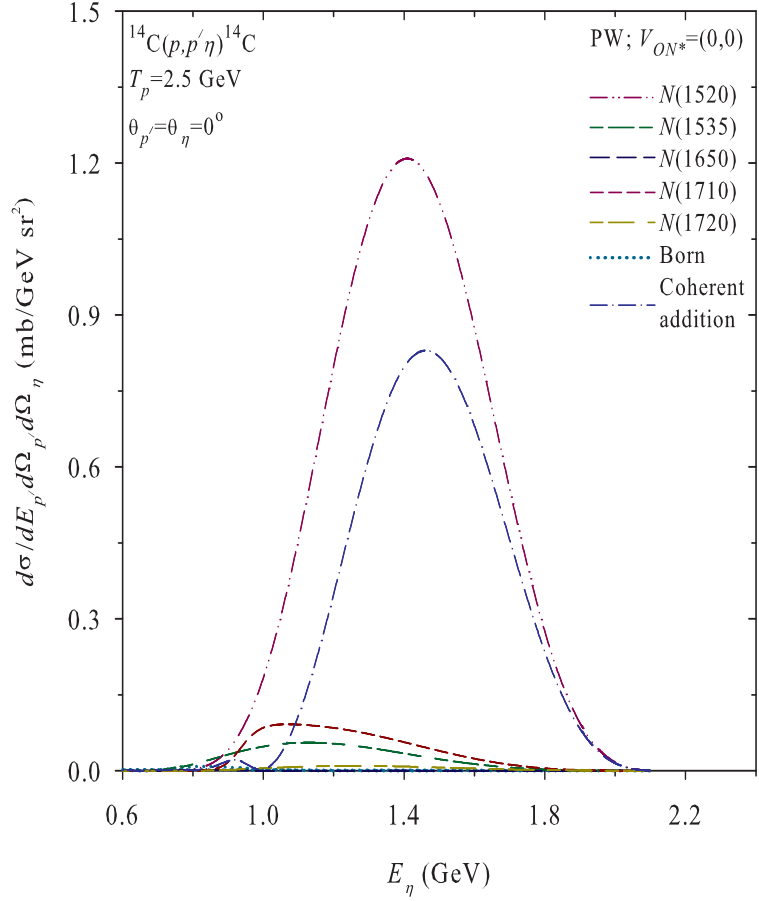,height=12.0 cm,width=10.0 cm}
}}
\caption{
(color online).
The $\eta$ meson energy $E_\eta$ distribution spectra for $^{14}$C
nucleus at 2.5 GeV. The cross section due to $N(1520)$ resonance is
distinctly largest (dot-dot-dash curve).
The coherently added cross section (dot-dash curve) is less than the
previous because of the interference of Born terms and resonances quoted
in the figure.
}
\label{fgrs}
\end{center}
\end{figure}

The calculated plane wave ($V_{ON^*}$ not included) results at 2.5 GeV
are illustrated in Fig.~\ref{fgrs}. The coherent contribution of $\pi^0$ and
$\eta$ meson exchange potentials, i.e., $V_{\pi^0}(q)$ and $V_\eta(q)$,
are incorporated in these results.
This
figure represents the cross sections due to Born terms, resonances
(mentioned earlier) and that occurring because of their coherent
contributions.
The
dot-dot-dash curve in this figure shows that the cross section because
of $N(1520)$ resonance is distinctly largest. Compared to it, the
cross sections due to Born terms and other resonances are insignificant.
The
interferences of Born terms and resonances in the coherently added
cross section, presented by dot-dash curve, is visible in the figure.
The peak of this cross section arises close to that because of $N(1520)$
resonance.

\begin{figure}[h]
\begin{center}
\centerline {\vbox {
\psfig{figure=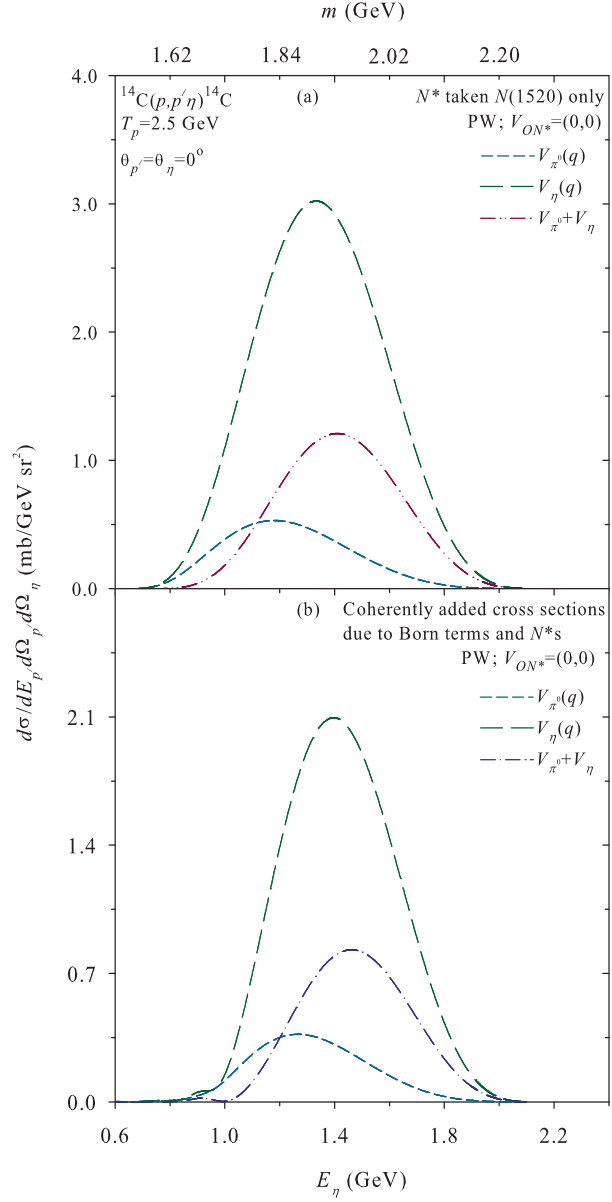,height=16.0 cm,width=08.0 cm}
}}
\caption{
(color online).
Upper part:
Contributions of $\pi^0$ and $\eta$ meson exchange potentials to the
cross section arising due to $N(1520)$ resonance only. The peak position
is shifted to the higher value of $E_\eta$ because of the interference of
these potentials (dot-dot-dash curve).
Lower part:
Same as those presented in the upper part but for the coherently added
cross sections due to Born terms and resonances (see text).
}
\label{fgpt}
\end{center}
\end{figure}

The cross sections of the considered reaction due to $V_{\pi^0}(q)$
and $V_\eta(q)$ at 2.5 GeV is described in Fig.~\ref{fgpt}. Along with
them, the coherently added cross section because of these potentials
is also presented in this figure.
The
upper part of it, i.e., Fig.~\ref{fgpt}(a), shows the calculated results
(plane wave; $V_{ON^*}$ not included) arising due to $N(1520)$ resonance
only, since the cross section because of this resonance (as shown in the
previous figure) is distinctly largest.
The
cross section due to $V_{\pi^0}$ (short-dash curve) is significantly
smaller $(\sim \frac{1}{5})$ than that due to $V_\eta$ (large-dash curve).
The
interference of these potentials is noticeable in the coherently
added cross section, see dot-dot-dash curve in this figure.

The smaller cross section arising because of $\pi^0$ meson exchange
potential $V_{\pi^0}$ over that due to $\eta$ meson exchange potential
$V_\eta$ may be understood, as an initial thought, by analyzing the
ratios of various factors appearing in
$ \left | \frac{V_{\pi^0}}{V_\eta} \right |^2 $ at the respective peak
of the cross sections.
The
coupling constants, quoted below Eqs.~(\ref{lag1}) and (\ref{lag3}), show
the ratio $ \left | \frac{ g_\pi f^*_\pi }{ g_\eta f^*_\eta} \right |^2 $
is approximately equal to 1.23.
The
ratio of isospin matrix elements in 
$ \left | \frac{V_{\pi^0}(q)}{V_\eta(q)} \right |^2 $ is $\frac{1}{49}$,
as the isospin contribution of a proton cancels that of a neutron in
the nucleus for $V_{\pi^0}(q)$.
Refering
to Fig.~\ref{fgpt}(a), the peak cross section due to $V_{\pi^0}(q)$
appears at the four-momentum transfer $q^2 \simeq -0.18$ GeV$^2$ where
as that because of $V_\eta(q)$ arises at $q^2 \approx -0.25$ GeV$^2$.
The
form factors at the respective peaks, according to Eq.~(\ref {ffc}), are
$ F_\pi(q^2 \simeq -0.18 ~\mbox{GeV}^2 )
= F^*_\pi(q^2 \simeq -0.18 ~\mbox{GeV}^2) \simeq 0.89 $
and $ F_\eta(q^2 \approx -0.25 ~\mbox{GeV}^2  )
= F^*_\eta(q^2 \approx -0.25 ~\mbox{GeV}^2  ) \simeq 0.78 $. These values
give
$ \left | \frac{ F_\pi(q^2) }{ F_\eta(q^2) } \right |^4 \simeq 1.7 $.
The
values of the pseudoscalar meson propagators are
$ G_\pi(q^2 \simeq -0.18 ~\mbox{GeV}^2) \simeq -5.06 $ GeV$^{-2}$ and
$ G_\eta(q^2 \approx -0.25 ~\mbox{GeV}^2) \approx -1.83 $ GeV$^{-2}$,
which show
$ \left | \frac{ G_\pi(q^2) }{ G_\eta(q^2) } \right |^2 \approx 7.65 $.
The
product of these factors shows
$ \left | \frac{V_{\pi^0}(q)}{V_\eta(q)} \right |^2
\simeq \frac{1}{3.06} $.
But,
the calculated results show the peak cross section due to $V_\eta$ is
$\sim 5$ larger than that because of $V_{\pi^0}$.

To resolve the above discrepancy (i.e., a factor of $\sim 1.63$), the
$N^*(m) \to N\eta$ decay probabilities (for $N^* \equiv N(1520)$) at
the peaks quoted in the previous analysis are considered.
It
is noticeable in Fig.~\ref{fgpt}(a) that the peak cross section because
of $V_{\pi^0}$ appears at $E_\eta \simeq 1.18$ GeV which corresponds to the
resonance mass $m \approx 1.83$ GeV, as mentioned on the upper x-axis of
this figure. The peak of the cross section because of $V_\eta$ appears at
$m \simeq 1.9$ GeV.
The
ratio of the decay probabilities at the respective peaks
$ \frac{ \Gamma_{N^* \to N\eta} (m \approx 1.83 ~\mbox{GeV})|_{V_{\pi^0}} }
{ \Gamma_{N^* \to N\eta} (m \simeq 1.9 ~\mbox{GeV})|_{V_\eta} } $,
according to Eq.~(\ref{wdth}), is close to $\frac{1}{1.61}$.
Therefore,
the above analyses justify the lesser cross section due to $V_{\pi^0}$
over that because of $V_\eta$, as visible in Fig.~\ref{fgpt}(a).

The importance of $V_{\pi^0}$ and $V_\eta$ in the coherently added
cross sections due to Born terms and quoted resonances are presented
in Fig.~\ref{fgpt}(b).
These
spectra show features which are qualitatively similar to those elucidated
in the upper part of this figure. This occurs since the distinctly
dominant cross section, as mentioned earlier, arises because of $N(1520)$
resonance.

To include the hadron nucleus interaction (optical potential) in the
calculated cross section, the distorted wave functions are used for $\chi$s
in $T$-matrices (given in Eqs~(\ref{tmxB}) and (\ref{tmxR})) and
$G_{N^*}$ in Eq.~(\ref{nsp0}) is replaced by the in-medium resonance
propagator.
Using
Glauber model \cite{glub, das05}, $\chi$ for beam proton $p$ can be
written as
\begin{equation}
\chi^{(+)} ({\bf k}_p, {\bf r}) = e^{i {\bf k}_p. {\bf r} }
exp[ -\frac{i}{v_p} \int^z_{-\infty} dz^\prime V_{Op} ({\bf b}, z^\prime) ].
\label{dwptn}
\end{equation}
For outgoing particles, i.e., $p^\prime$ and $\eta$ meson, the form for
the distorted wave functions is 
\begin{equation}
\chi^{(-)^*} ({\bf k}_X, {\bf r}) = e^{-i {\bf k}_X. {\bf r} }
exp[ -\frac{i}{v_X} \int^{+\infty}_z dz^\prime V_{OX} ({\bf b}, z^\prime) ];
~~~~~ (X = p^\prime, \eta).
\label{dwpet}
\end{equation}
$v$ and $V ({\bf b}, z^\prime)$ in above equations represent the velocity
and optical potential respectively of the continuum particle. These
potentials describe the initial and final state interactions of the
reaction.

Incorporating the resonance nucleus interaction $V_{ON^*}({\bf r})$ in
$G_{N^*}(m)$ given in Eq.~(\ref{nsp0}), the resonance propagator in the
nucleus can be expressed as
\begin{equation}
G_{N^*} (m, {\bf r}) = \frac{1}
{m^2-m^2_{N^*} + im_{N^*}\Gamma_{N^*}(m) - 2E_{N^*}V_{ON^*}({\bf r})},
\label{nspr}
\end{equation}
where $E_{N^*}$ is the energy of the resonance $N^*$.

The optical potential $V_{OX} ({\bf r})$, appearring in Eqs.~(\ref {dwptn})
- (\ref {nspr}), is calculated using $ ``t\varrho ({\bf r})" $
approximation \cite{das05}, i.e.,
\begin{equation}
V_{OX} ({\bf r})
= -\frac{v_X}{2} [i+\alpha_{XN}] \sigma^{XN}_t \varrho ({\bf r}),
\label{opts}
\end{equation}
where the symbol $X$ represents either a proton or a resonance $N^*$.
$v_X$ is the velocity of the particle $X$.
$\alpha_{XN}$
denotes the ratio of the real to imaginary part of $X$-nucleon scattering
amplitude $f_{XN}$, and $\sigma^{XN}_t$ is the corresponding total
cross section.
To
evaluate the proton nucleus optical potential, i.e., $ V_{Op} ({\bf r}) $
as well as $ V_{Op^\prime} ({\bf r}) $, the energy dependent
experimentally determined values for $\alpha_{pN}$ and $\sigma^{pN}_t$
have been used \cite{nndt}.
The
measured values for $N^*$-nucleon scattering parameters, i.e.,
$\alpha_{N^*N}$ and $\sigma^{N^*N}_t$, are not available. To estimate them,
$ \alpha_{N^*N} \approx \alpha_{pN} $ and $ \sigma^{N^*N}_{el} 
\approx \sigma^{pN}_{el} $ are taken since the elastic scattering dynamics
of $N^*$ can be assumed not much different from that of a
proton \cite{londergan}.
For
the reactive part of $\sigma^{N^*N}_t$, the dynamics of $N^*$ can be
considered same as that of a nucleon at its kinetic energy enhanced by
$\Delta m$,
i.e.,
$ \sigma_r^{N^*N} (T_{N^*N}) \approx \sigma_r^{NN} (T_{N^*N}+\Delta m) $.
Here, $\Delta m$ is the mass difference between the resonance and
nucleon. $T_{N^*N}$ is the total kinetic energy in the $N^*N$ center
of mass system \cite{londergan}.

The $\eta$ meson optical potential $V_{O\eta} ({\bf r})$ is evaluated
from its self-energy $\Pi_\eta ({\bf r})$ in the nucleus.
The resonance hole contribution to $\Pi_\eta ({\bf r})$, according to
Alvaredo and Oset \cite{oset}, can be written as
\begin{eqnarray}
\Pi_\eta ({\bf r}) = 2E_\eta V_{O\eta} ({\bf r}) = \sum_{N^*} |C(N^*)|^2
\frac{ \varrho ({\bf r}) }
{ m - m_{N^*} + \frac{i}{2}\Gamma_{N^*}(m) - V_{ON^*} ({\bf r})
                                            + V_{ON} ({\bf r}) }.
\label{opet}
\end{eqnarray}
The prefactor $|C(N^*)|^2$ in this equation depends on $N^*$ used to
evaluate $\Pi_\eta ({\bf r})$. The nucleon potential energy in the nucleus
is taken as
$ V_{ON} ({\bf r}) = -50 \varrho ({\bf r}) / \varrho (0) $ MeV
\cite{oset}. $\Pi_\eta ({\bf r})$ arising due to nucleon hole pair is
worked out following that due to $\pi^0$ meson, see page 157 in
Ref.~\cite{erwe0}.

\begin{figure}[h]
\begin{center}
\centerline {\vbox {
\psfig{figure=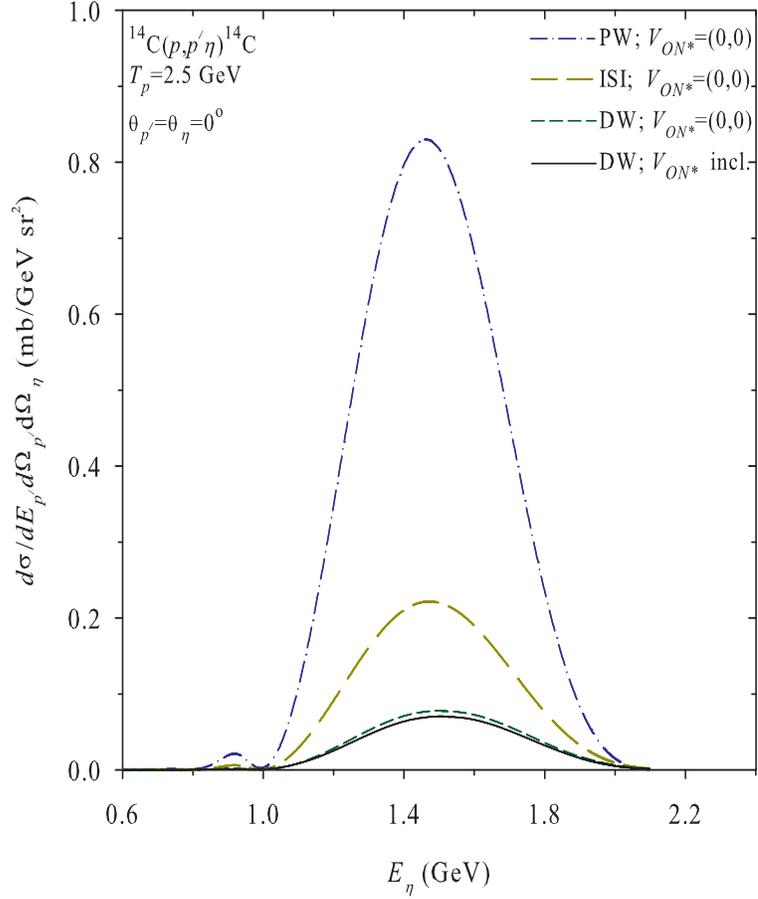,height=12.0 cm,width=10.0 cm}
}}
\caption{
(color online).
The sensitivity of the cross section to the hadron nucleus interaction
(optical potential). The cross section is reduced drastically (by a
factor of $\simeq 12$), and the peak position is shifted by 40 MeV towards
the higher value of $E_\eta$ because of these interactions.
}
\label{fgit}
\end{center}
\end{figure}

The sensitivity of the calculated cross section to the hadron nucleus
interaction is exhibited in Fig.~\ref{fgit}. The large peak (shown in this
figure) in the coherently added cross sections (arising because of Born
terms and considered resonances as well as because of $\pi^0$ and $\eta$
meson exchange potentials) is considered for this purpose.
The
dot-dash curve represents the plane wave ($V_{ON^*}$ not included)
cross section of the considered reaction (also shown earlier).
The
cross section is reduced by a factor of 3.74 because of the
incorporation of initial state interaction (ISI), see the long-dash
curve.
The
short-dash curve elucidates the calculated spectrum obtained after the
inclusion of both ISI and FSI (final state interaction), i.e., it
describes the distorted wave results where $V_{ON^*}$ is not taken into
consideration. The cross section is further reduced by a factor of 2.86
due to the inclusion of FSI.
The
solid curve represents the calculated distorted wave results where
$V_{ON^*}$ has been incorporated. It shows the change in the cross section
due to this potential is negligible.
Therefore,
the calculated plane wave ($V_{ON^*}$ not included) cross section is
reduced drastically, i.e., by a factor of 11.81, because of the
inclusion of all hadron nucleus interactions.
The
shift in the peak position because of these interactions is about 40
MeV towards the higher value of $E_\eta$ in the spectrum.

\begin{figure}[h]
\begin{center}
\centerline {\vbox {
\psfig{figure=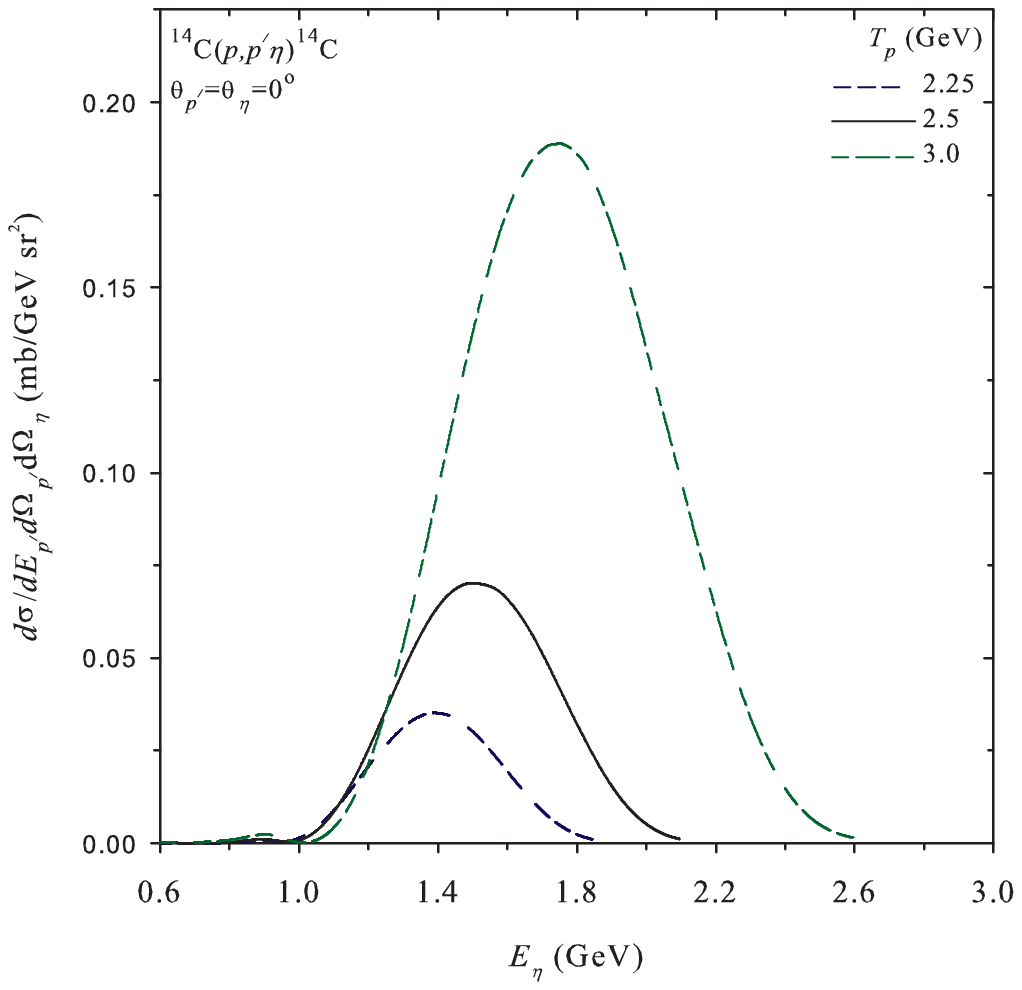,height=12.0 cm,width=10.0 cm}
}}
\caption{
(color online).
Beam energy dependance of the cross section. The cross section increases
and the peak position shifts towards the higher $E_\eta$ with the
enhancement in the beam energy.
}
\label{fgbm}
\end{center}
\end{figure}

The beam energy dependence of the distorted wave results are elucidated
in Fig.~\ref{fgbm}. The resonance nucleus interactions are also included
in this results.
The
cross section at the large peak is increased by a factor of 5.37, and
the peak position is shifted from $E_\eta \simeq 1.39$ GeV to
$E_\eta \simeq 1.75$ GeV with the increase in the beam energy from
2.25 GeV to 3 GeV.

\section{Conclusions}

The differential cross sections have been calculated for the coherent
$\eta$ meson energy $E_\eta$ distribution in the proton induced reaction
on the scalar-isovector nucleus.
The
$\eta$ meson in the final state is considered to arise because of Born
terms and resonances produced in the intermediate state.
The
interaction between the projectile proton and a nucleon in the target
nucleus is described by the pseudoscalar meson, i.e., $\pi^0$ and $\eta$
mesons, exchange potentials.
The
calculated results show that the distinctly dominant contribution to the
$\eta$ meson production cross section arises because of $N(1520)$
resonance. The coherently added cross section arising due to Born terms
and considered resonances is less than the previous due to their
interferences.
The
cross section because of $\pi^0$ meson exchange potential is lesser than
that due to $\eta$ meson exchange potential.
The
interference of these potentials is distinctly visible in the $\eta$
meson energy $E_\eta$ distribution spectrum.
The
cross section is reduced drastically and shifted towards higher
$E_\eta$ because of the hadron nucleus interactions, i.e., initial and
final state interactions including resonance nucleus interactions.
The
calculated results show that the cross section is very sensitive to the
beam energy, as the magnitude of cross section increases and its peak
position shifts towards higher $E_\eta$ with the enhancement in the
beam energy.

\section{Acknowledgement}

The author thanks the anonymous referee for giving the comments which
improved the quality of this work. Dr. L. M. Pant is acknowledged for his
help in editing the manuscript.

\newpage

{\bf Figure Captions}

\begin{enumerate}

\item
(color online).
Schematic diagrams describing the mechanism of the considered reaction
(see text): (a) direct (or post-emission) and (b) cross (or pre-emission)
diagrams.

\item
(color online).
The $\eta$ meson energy $E_\eta$ distribution spectra for $^{14}$C
nucleus at 2.5 GeV. The cross section due to $N(1520)$ resonance is
distinctly largest (dot-dot-dash curve).
The coherently added cross section (dot-dash curve) is less than the
previous because of the interference of Born terms and resonances quoted
in the figure.

\item
(color online).
Upper part:
Contributions of $\pi^0$ and $\eta$ meson exchange potentials to the
cross section arising due to $N(1520)$ resonance only. The peak position
is shifted to the higher value of $E_\eta$ because of the interference of
these potentials (dot-dot-dash curve).
Lower part:
Same as those presented in the upper part but for the coherently added
cross sections due to Born terms and resonances (see text).

\item
(color online).
The sensitivity of the cross section to the hadron nucleus interaction
(optical potential). The cross section is reduced drastically (by a
factor of $\simeq 12$), and the peak position is shifted by 40 MeV towards
the higher value of $E_\eta$ because of these interactions.

\item
(color online).
Beam energy dependence of the cross section. The cross section increases
and the peak position shifts towards the higher $E_\eta$ with the
enhancement in the beam energy.

\end{enumerate}

\newpage

\newpage

\newpage

\newpage

\newpage


\newpage
\begin{thebibliography}{100}

\bibitem{wgpr}
V. Baru et al., arXiv/nucl-th:0610011.

\bibitem{cosy}
S. Schadmand (for WASA at COSY Collaboration), Pramana {\bf 75} (2010) 225.

\bibitem{satu}
J. Berger et al., Phys. Rev. Lett. {\bf 61} (1988) 919;
R. Frascaria et al., Phys. Rev. C {\bf 50} (1994) R537.

\bibitem{loal}
J. C. Peng et al., Phys. Rev. Lett. {\bf 63} (1989) 2353.

\bibitem{broo}
W. B. Tippens et al., Phys. Rev. D {\bf 63} (2001) 052001.

\bibitem{gsi}
F. -D. Berg et al., Phys. Rev. Lett. {\bf 72} (1994) 977.

\bibitem{peeta}
M. Williams et al., Phys. Rev. C {\bf 80} (2009) 045213;
B. Krusche et al., Phys. Rev. Lett. {\bf 74} (1995) 3736;
J. W. Price et al., Phys. Rev. C {\bf 51} (1995) 2283(R);
B. Krusche et al., Phys. Lett. B {\bf 358} (1995) 40;
M. R$\ddot{\mbox{o}}$big-Landau et al.,
Phys. Lett. B {\bf 373} (1996) 45.

\bibitem{etmn}
R. S. Bhalerao and L. -C. Liu, Phys. Rev. Lett. {\bf 54} (1985) 865;
Q. Haider and L. -C. Liu, Phys. Lett. B {\bf 172} (1986) 257;
L. -C. Liu and Q. Haider, Phys. Rev. C {\bf 34} (1986) 1845;
G. L. Li, W. K. Cheng and T. T. S. Kuo, Phys. Lett. B {\bf 195} (1987) 515.

\bibitem{cosy2}
R. E. Chrien et al., Phys. Rev. Lett. {\bf 60} (1988) 2595;
A. Budzanowski et al., (COSY-GEM Collaboration),
Phys. Rev. C {\bf 79} (2009) 012201(R).

\bibitem{pemx}
C. Y. Cheung, Phys. Lett. B {\bf 119} (1982) 47; 
C. Wilkin, Phys. Lett. B {\bf 331} (1994) 276.

\bibitem{mosl}
J. Lehr, M. Post and U. Mosel, Phys. Rev. C {\bf 68} (2003) 044601;
J. Lehr and U. Mosel, ibid, 044603; R. C. Carrasco, ibid, {\bf 48} (1993)
2333.

\bibitem{shrf}
M. Hedayati-Poor and H. S. Sherif, Phys. Rev. C {\bf 76} (2007) 055207;
M. Hedayati-Poor, S. Bayegan and H. S. Sherif,
Phys. Rev. C {\bf 68} (2003) 045205.

\bibitem{petp}
Ch. Sauerman, B. L. Friman and W. N$\ddot{\mbox{o}}$renberg,
Phys. Lett. B {\bf 341} (1995) 261;
J. -F. Germond and C. Wilkin, Nucl. Phys. A {\bf 518} (1990) 308;
A. Moalem, E. Gedalin, L. Razdolskaja and Z. Shorer,
Nucl. Phys. A {\bf 589} (1995) 649.

\bibitem{lawl}
J. M. Laget, F. Wellers and J. F. Lecolley,
Phys. Lett. B {\bf 257} (1991) 254.

\bibitem{vett}
T. Vetter, A Engel, T. Bri$\acute{\mbox{o}}$ and U. Mosel,
Phys. Lett. B {\bf 263} (1991) 153.

\bibitem{oset}
B. Lo$\acute{\mbox{p}}$ez Alvaredo and E. Oset,
Phys. Lett. B {\bf 324} (1994) 125.

\bibitem{das14}
Swapan Das, Phys. Lett. B {\bf 737} (2014) 75.

\bibitem{pdg12}
J. Beringer et al., Particle Data Group,
Phys. Rev. D  {\bf 86} (2012) 010001.

\bibitem{liu89}
L. -C. Liu, J. T. Londergan, and G. E. Walker,
Phys. Rev. C {\bf 40} (1989) 832.

\bibitem{peters}
W. Peters, H. Lenske and U. Mosel, Nucl. Phys. A {\bf 642} (1998) 506.

\bibitem{blokland}
I. R. Blokland and H. S. Sherif, Nucl. Phys. A {\bf 694} (2001) 337.

\bibitem{erwe0}
T. Ericson and W. Weise, Pions and Nuclei
(Clarendon Press, Oxford, 1988), p.~18.

\bibitem{chiang}
H. C. Chiang, E. Oset and L. -C. Liu, Phys. Rev. C {\bf 44} (1991) 738;
R. Machleidt, K. Holinde and Ch. Elster, Phys. Rep. {\bf 149} (1987) 1.

\bibitem{das04}
Swapan Das, Phys. Rev. C {\bf 70} (2004) 034604.

\bibitem{fadlt}
G. F$\ddot{\mbox{a}}$ldt and C. Wilkin, Nucl. Phys. A {\bf 587} (1995) 769;
A. Fix and H. Arenh$\ddot{\mbox{o}}$vel, Nucl. Phys. A {\bf 620} (1997) 457.

\bibitem{manley}
D. M. Manley, R. A. Arndt, Y. Goradia and V. I. Teplitz,
Phys. Rev. D {\bf 30} (1984) 904;
D. M. Manley, Phys. Rev. D {\bf 51} (1995) 4837; ibid,
Int. J. Mod. Phys. A {\bf 18} (2003) 441.

\bibitem{andt}
C. W. De Jager, H. De. Vries and C. De Vries,
At. Data Nucl. Data Tables, {\bf 14} (1974) 479.

\bibitem{glub}
R. J. Glauber, in Lectures in theoretical physics, vol.~1, edited by
W. E. Brittin et al., (Interscience Publishers, New York, 1959).

\bibitem{das05}
Swapan Das, Phys. Rev. C {\bf 72} (2005) 064619.

\bibitem{nndt}
D. V. Bugg, et al., Phys. Rev. {\bf 146} (1966) 980;
S. Barshay, et al., Phys. Rev. C {\bf 11} (1975) 360;
W. Grein, Nucl. Phys. B {\bf 131} (1977) 255;
C. Lechanoine-Leluc and F. Lehar, Rev. Mod. Phys. {\bf 65} (1993) 47;
R. M. Barnett et al., Particle Data Group,
Phys. Rev. D {\bf 54}, (1996) 192; http://pdg.lbl.gov/xsect/contents.html.

\bibitem{londergan}
B. K. Jain, N. G. Kelkar and J. T. Londergan,
Phys. Rev. C {\bf 47} (1993) 1701.


\end{thebibliography}
\end{document}